\documentclass[conference]{IEEEtran}
\IEEEoverridecommandlockouts
\usepackage{cite}
\usepackage{amsmath,amssymb,amsfonts}
\usepackage{graphicx}
\usepackage{textcomp}
\usepackage{xcolor}
\usepackage{hhline}
\usepackage{algorithm}
\usepackage{algpseudocode}
\usepackage{bm}

\def\BibTeX{{\rm B\kern-.05em{\sc i\kern-.025em b}\kern-.08em
    T\kern-.1667em\lower.7ex\hbox{E}\kern-.125emX}}
\begin{document}

\title{Detection of Violent Extremists in Social Media\\
%{\footnotesize \textsuperscript{*}Note: Sub-titles are not captured in Xplore and
%should not be used}
%\thanks{Identify applicable funding agency here. If none, delete this.}
}

\author{\IEEEauthorblockN{Hamidreza Alvari, Soumajyoti Sarkar, Paulo Shakarian}
%\IEEEauthorblockA{%\textit{dept. name of organization (of Aff.)} \\
\textit{Arizona State University}\\
Tempe, USA \\
\{halvari, ssarka18, shak\}@asu.edu}
%}

\maketitle

\begin{abstract}
The ease of use of the Internet has enabled violent extremists such as the Islamic State of Iraq and Syria (ISIS) to easily reach large audience, build personal relationships and increase recruitment. Social media are primarily based on the reports they receive from their own users to mitigate the problem. Despite efforts of social media in suspending many accounts, this solution is not guaranteed to be effective, because not all extremists are caught this way, or they can simply return with another account or migrate to other social networks. In this paper, we design an automatic detection scheme that using as little as three groups of information related to usernames, profile, and textual content of users, determines whether or not a given username belongs to an extremist user. We first demonstrate that extremists are inclined to adopt usernames that are similar to the ones that their like-minded have adopted in the past. We then propose a detection framework that deploys features which are highly indicative of potential online extremism. Results on a real-world ISIS-related dataset from Twitter demonstrate the effectiveness of the methodology in identifying extremist users.
\end{abstract}


\section{Introduction}
Last years have witnessed a huge rise in the threat of radicalized extremists groups who seek to commit insurgent attacks around the globe. While new technologies such as the Internet and social media are now being used by many users~\cite{beigi2018similar,alvari2011community}, they have also been leveraged extensively by radicalized groups to make relationships with their audience and recruit new members~\cite{isi10}. Social media platforms such as Twitter are now being utilized by terrorist organizations, to directly communicate with their worldwide audience~\cite{informs16}. The free and unregulated nature of these tools, have helped extremists to easily form online communities and disseminate their beliefs and training materials, without a fear of getting censored on traditional media outlets.

Recently, social networks have begun to actively fight against these groups. In august 2016, Twitter which has been long believed to be the main propaganda for ISIS, finally took serious actions by shutting down over 36,000 ISIS related accounts~\cite{twitter16}. Social media are primarily based on the reports they receive from their own users, or from specific teams assigned to mitigate the problem. Recently they began to explore more effective ways such as using algorithms (e.g. spam-fighting algorithms) to automatically detect any violent content as supplements to these reports and boost performance~\cite{si14}. Despite the huge effort of these social networks in shutting down many accounts, this solution is not guaranteed to be effective, because not all extremists are caught this way, and the owners of suspended accounts can simply return with another account or migrate to other social networks. Consequently, extra efforts need to be dedicated to proposing capabilities that could be deployed by authorities to combat radicalized extremists and mitigate their threats, regardless of the underlying social network platform. 

\textbf{Present work. }In this paper, we design a detection scheme that using as little as three groups of information (inspired from the literature~\cite{10bits,informs16,socinfo16}), can determine whether or not a given username\footnote{In this work, we may use the terms usernames and handles interchangeably} belongs to an extremist user. Specifically, we use a dataset from Twitter~\cite{alvari2018early} and first show that \textit{extremist users on Twitter tend to adopt handles that follow similar patterns, in contrast to the normal users}. Then, a detection framework is proposed to identify if a given Twitter handle belongs to an extremist given its proximity to an existing set of extremist-related handles. We compare different supervised and semi-supervised approaches using the features from Twitter handle, profile information and content which are highly indicative of online extremism. To further show the significance of the features we conduct significance analysis of the features using the labeled instances and feature selection measure $\chi^2$ and compare our results against char-LSTM which automatically extracts features. 

\textbf{Contributions. }Our main contributions are thus summarized in below:
\begin{itemize}	  
	\item We first demonstrate that extremists on Twitter are inclined towards adopting handles with similar patterns.To that end, we used the well-known Lavenshtein ratio as a measure of distance between two Twitter handles and performed two-sample t-test to demonstrate that compared to normal users, extremists tend to choose similar handles.  
	\item We propose three main groups of features, related to the Twitter handles, profile information and tweet-level content. Overall, our feature engineering scheme has 13 features which are then fed into different supervised and semi-supervised learners.
	\item Results on a real-world ISIS-related dataset demonstrate that the introduced features are effective in detecting online violent extremists in social media. 	
\end{itemize}

\textbf{Observations. }We make the following observations:

\begin{itemize}
	\item The highest precision of 0.96 in identification of the extremists belongs to SVM. This is in line with the previous research that SVM performs well on textual data.
	\item Among several approaches used in this work, char-LSTM and the semi-supervised approach LabelSpreading with RBF kernel achieve the equal and highest F1-score of 0.76. The fact that the LabelSpreading achieves comparable performance as char-LSTM, further demonstrates the effectiveness of the proposed feature engineering scheme, as LSTM has shown promising results in the literature while it does not use any hand-crafted features.  
	\item Char-LSTM achieves a precision of 0.77 while maintaining a high recall of 0.76 on the positive class. This suggests that the memory module in LSTM can help in minimizing the number of false negatives.
\end{itemize}

%The remainder of the paper is organized as follows. In Section 2, we explain the dataset used in this work. Section 3 discusses our methodology for detection of online extremists. In Section 4, we discuss the experiments. Section 5 reviews the related work. We conclude the paper in Section 6 and discuss the future work.

\section{Related Work}
The explosive growth of the Web has raised numerous security and privacy issues. Mitigating these concerns has been studied from different aspects~\cite{beigi2019untraceable,alvari2016isi,beigi2018privacy,Beigi2018HT}. For instance, several studies have focused on understanding extremism in social networks~\cite{benigni2017online,Benigni2016,informs16,socinfo16,si14,transac15,isi16}. For instance, the work of~\cite{informs16}, uses Twitter and proposes an approach to predict new extremists. They also determine if the newly created account belongs to a suspended extremist, and predict the ego-network of the suspended extremist upon creating her new account. Their approach integrates variants of the logistic regression with optimized search policies to detect the new accounts of returning suspended extremist users. They (1) use potential features of an account to predict if this account belongs to an extremist user, (2) determine if multiple accounts belong to the same extremist user, based on the fact that new account shall resemble the suspended account in different aspects, (3) predict whom the suspended extremist user is most likely to follow again, and finally (4) develop a network search policy to find the suspected users upon returning to a social network. Similar work of~\cite{socinfo16} uses tweets to build models to predict (1) if a pro-ISIS user's account will be suspended due to the extremist content, (2) which users will adopt and retweet ISIS content, and (3) which users will have interactions with pro-ISIS users. To do so, the authors use logistic regression and random forest classifiers for different types of prediction tasks. They deploy variety of features across different dimensions, such as user meta-data, network statistics and temporal patterns of activity. Two scenarios are then designed for each prediction task: a time independent (static) one which does not take into account the temporal dependencies, and a simulated real-time one by considering the timeline of content availability. The difference between these two studies is, authors of~\cite{informs16} also study other aspects including identifying multiple accounts for an extremist user, re-following suspended accounts' connections and searching for the suspended extremist users who might return to a social media.

Other works also seek to identify the extremist content in radicalized groups beyond ISIS. The work of~\cite{si14} uses data from Jihadist website Ansar AlJihad Network to develop supervised learning and NLP techniques to automatically detect cyber-recruitment of extremist groups. A comparison is done between classifiers na\"ive Bayes, logistic regression, classification trees, boosting and SVM, for labeling forum posts as either related or not related to the recruitment of extremist groups. They leverage the bag-of-words technique to convert the corpus into a term-document matrix, following the standard routine of the preprocessing techniques such as basic normalization and stemming. Similarly in~\cite{transac15}, same Jihadist network along with their previously developed SVM classifier are used to automatically identify recruitment posts. Their previous work shall be served as pre-screening step to reduce the efforts made by human analysts to manually hand-label the documents. In their new study, the textual content of the dataset is analyzed with latent Dirichlet allocation (LDA) and fed into several time-series models to predict cyber-recruitment. This new research conducted by the same authors complemented their previous study, by applying current natural language processing and time series analysis techniques to forecast the recruitments.

Beyond these works, the work of~\cite{isi16} takes a different approach to track individual's behavioral indicators of homegrown extremism, using public and law enforcement data. The intuition is to use graph pattern matching to identify suspicious trajectories and potential radicalization over a dynamic heterogeneous graph associated with the fused data from public and law enforcement. The authors first develop a query pattern of radicalization and then run several graph pattern matching algorithms to detect and track the on-going radicalization. They develop the investigative simulation graph pattern matching technique, which is composed of required extension to the existing dual simulation graph pattern matching method to avoid over-matching. This approach provides analysts and law enforcement officials with the ability to find partial/full matches, given a query of radicalization, as well as the pace of the appearance of the radicalized extremists. As opposed to the above studies, in this paper, we make the first attempt on determining if a given Twitter handle belongs to an extremist user or not, using only little information gathered from the handle, profile and content.

\section{Data Preparation}
%In this work we use two datasets for evaluating the proposed framework. 

%\textbf{Dataset 1.} This dataset contains 2.7K Twitter accounts who were suspended by Twitter because of the connections to Russia's Internet Agency troll farm. The list\footnote{https://democrats-intelligence.house.gov/uploadedfiles/exhibit\_b.pdf} was released by the U.S. Congress as part of their investigation of the alleged Russian interference in the 2016 U.S. presidential election.  

The dataset was collected from Twitter and consists of approximately 1.6M tweets that were posted using 25 extremism-related hashtags such as \textsf{\#AbuBakralBaghdadi}, \textsf{\#ISIL}, \textsf{\#ISIS}, \textsf{\#Daesh}, and \textsf{\#IslamicState}. We construct our extremist labels (positive labels) by collecting a limited number of 150 suspended ISIS-related Twitter handles which were reported to the Twitter Safety account (@TwitterSafety) by normal users. To make a balanced labeled dataset, 150 random handles corresponding to normal users were also collected to serve as our negative labels. 

Inspired by the literature~\cite{10bits,informs16,socinfo16}, we define 3 major groups of overall 13 features, which could be leveraged to filter out less likely extremists. This way, we obtain 300K highly extremism related tweets from which we randomly pick a smaller sample of 3K handles who posted the tweets. The description of the dataset is shown in Table~\ref{tb:st}.

%This dataset contains ISIS related profiles, and was gathered from Feb 22, 2016 to May 27, 2016, using the Twitter streaming API\footnote{https://developer.twitter.com/en/docs} which provides 1\% random tweets from the total volume of tweets at a particular moment. The dataset was collected using 290 different hashtags such as \textsf{\#Terrorism}, \textsf{\#StateOfTheIslamicCaliphate}, \textsf{\#Rebels}, \textsf{\#BurqaState}. We use a subset of this dataset which contains ....

%After the data collection, we check through Twitter API to see whether each account has been suspended (PSM) or not (normal)~\cite{thomas2011suspended}. In this dataset 11\% of the users are PSM and the rests are normal.

\begin{table}
	\small
	\centering
	\caption{Description of the dataset.}
	\begin{tabular}{|l|c|c|}
		\cline{1-3}
		\textbf{Name}          & \multicolumn{2}{c|}{\textbf{Value}}\\
		\hhline{===}
		Raw      & \multicolumn{2}{c|}{1.6M}  \\ \cline{1-3}
		Filtered & \multicolumn{2}{c|}{300K} \\ \cline{1-3}
		Unlabeled (sampled) & \multicolumn{2}{c|}{3K} \\ \cline{1-3}
		Labeled  & \textbf{Positive} & \textbf{Negative} \\ \cline{2-3}
		& 150  & 150 \\
		\cline{1-3}
	\end{tabular}
	\label{tb:st}
\end{table}

\section{Methodology}
Here, we first present the introduced feature groups used to filter out less likely extremists from the data. Next, we will pose our research questions and seek to answer them.
\subsection{Feature Engineering}
\begin{figure}[t]
	\caption{2-D projection of the sampled filtered data using t-SNE transformation. Clustering techniques such as K-means will have difficulty to correctly assign labels to the unlabeled instances using only few existing labeled samples.}
	\centering
	\includegraphics[width=0.5\textwidth]{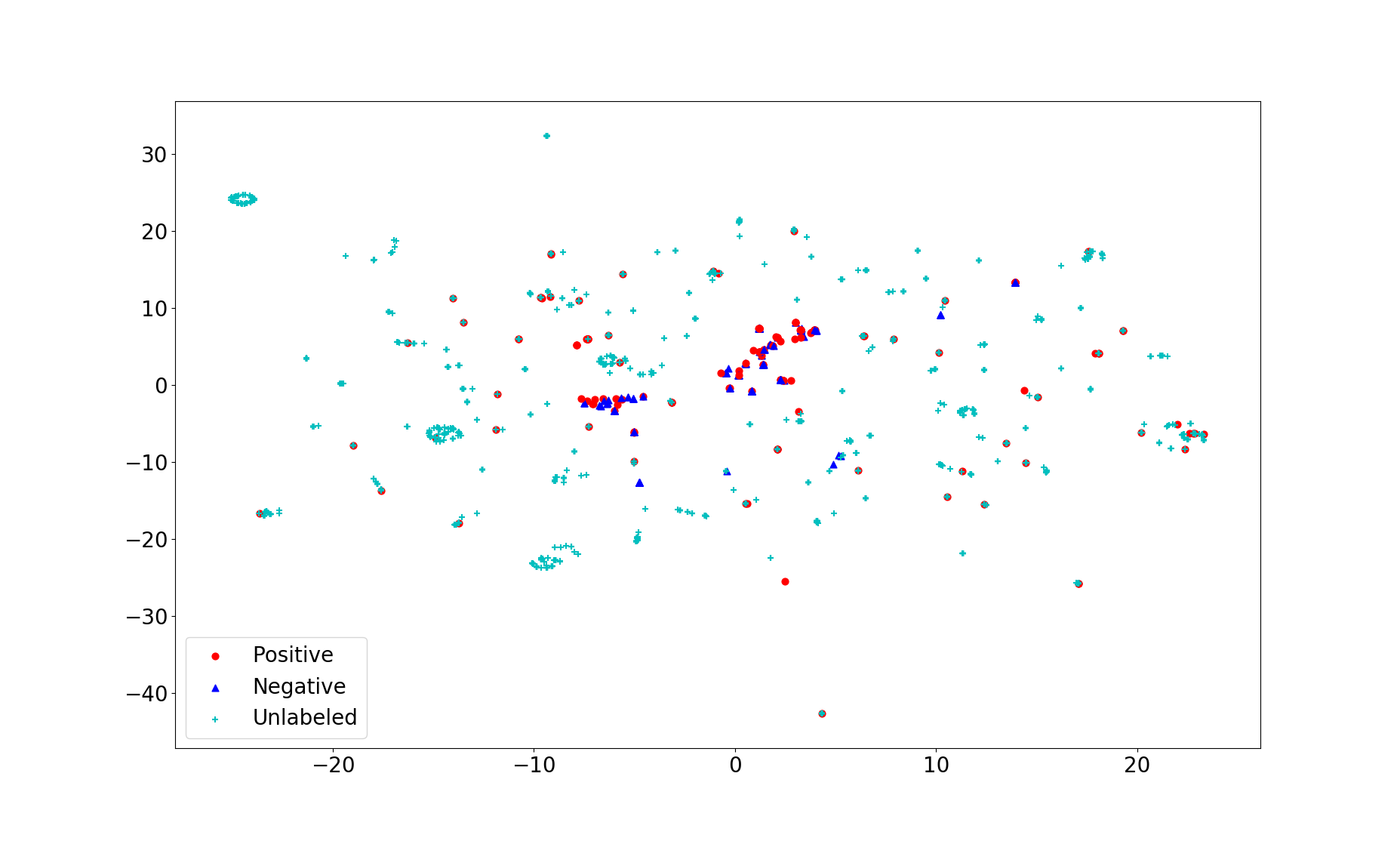}
	\label{fig:viz}
	\vspace{-5mm}
\end{figure}

We categorize the features used in this work into the following 3 major groups:
\begin{enumerate}
	\item \textbf{Twitter handle's related features}: this group contains 3 features related to the given handle, namely, length of the handle, number of unique characters in the handle, and complexity of the handle. To compute the complexity, Kolmogorov complexity is used, which is defined as the length of the shortest program to reproduce the handle on a universal machine such as Turing machine~\cite{Li:2008:IKC:1478784}. Since Kolmogorov complexity is computationally intractable, we use the Entropy of the handle as a way to approximate its complexity.
	
	\item \textbf{Profile related features}: this group contains 7 features related to the profile of the user who posted the tweet, including the number of her followers, friends and tweets, the existence of profile's description and location. Also, for the last two features in this group, we check if the account is verified and geo-enabled.
	
	\item \textbf{Content related features}: we have the following 3 features related to the content of the given tweet: the number of URLs, the number of hashtags and the sentiment of the content. For the sentiment, we check if the content has a higher negative score than its positive score. 
\end{enumerate}
For the sake of visualization, a 2-D projection of the sample of the filtered dataset (using t-SNE transformation~\cite{tsne}) is depicted in Fig.~\ref{fig:viz}. As it is seen, basic clustering techniques such as K-means will have difficulty to correctly assign labels to the unlabeled instances using only few existing labeled samples.

\subsection{Research Questions}
Having defined the feature engineering scheme in the previous section, here, we seek to answer the following research questions that will ultimately help identifying violent extremists in social media:

\begin{itemize}
	\item \textbf{RQ1: Are extremists on Twitter inclined to adopt similar handles?}
	\item \textbf{RQ2: Can we infer the labels (extremist vs. non-extremist) of unseen handles based on their proximity to the labeled instances?}
\end{itemize}
To answer the \textbf{first} question, for each pair of extremist users $(i,j)$ we compute the similarity between their corresponding handles $\langle s_i,s_j\rangle$ as follows. 
\begin{equation}
Sim(s_i,s_j) = \frac{1 - L(s_i,s_j)}{max(len(s_i), len(s_j))}
\end{equation}
where $L(s_i,s_j)$ is the well-known Lavenshtein ratio~\cite{lavenshtein} which is used as a measure of distance between the two strings $s_1$ and $s_2$. We create a vector $v_e$ whose elements are similarity scores between each pair of extremists. We repeat the procedure for each pair of extremist $i$ and a normal user $k$ and create a vector $v_{en}$. We conduct a two-sample t-test on $v_e$ and $v_{en}$ with the null and alternative hypotheses defined as $H_0:v_e\leq v_{en},~~ H_1:v_e > v_{en}$. The null hypothesis is rejected at significance level $\alpha=0.01$ and p-value of $p=0.009$, suggesting that extremists are biased towards adopting similar handles. Although this might seem a bit simplistic at the first sight, it has not been examined in the literature. Later, we will see how effective this simple idea could be in inferring the labels for unseen handles and help detecting the extremists by merely glancing at their handles.  

To answer the \textbf{second} question, let us first obtain our feature spaces associated with the labeled and unlabeled instances, by converting each handle to a vector of 5 features. We use the following features: \textit{length of the handle}, \textit{maximum number of occurrence of a character in the handle}, \textit{number of unique characters in the handle}, \textit{number of digits that the handle starts with}, and \textit{complexity of the handle}. Ultimately, these two feature spaces are fed as the inputs to the semi-supervised and supervised learners.

\begin{table}[t]\small
	\centering
	\caption{Comparison of the methods on the labeled data, for the positive (extremist) class.}%\vspace{-2mm}
	\begin{tabular}{|l|c|c|c|} 
		\cline{1-4}
		\textbf{Learner} & \textbf{Precision} &
		\textbf{Recall} &
		\textbf{F1-score}
		\\
		\hhline{====}
		SVM & \textbf{0.96} & 0.5 & 0.65 \\ \hline
		Char-LSTM & 0.77 & \textbf{0.76} & \textbf{0.76} \\ \hline
		LabelSpreading (RBF) & 0.85 & 0.69 & \textbf{0.76}\\ \hline
		Laplacian SVM & 0.89 & 0.6 & 0.7 \\ \hline
		LabelSpreading (KNN) & 0.83 & 0.67 & 0.73  \\ \hline
		Co-Training (SVM) & 0.9 & 0.53 & 0.66 \\ \hline
		KNN & 0.81 & 0.7 & 0.74 \\ \hline
		Gaussian NB & 0.89 & 0.56 & 0.69 \\ \hline
		Logistic Regression & 0.76 & 0.61 & 0.67 \\ \hline
		AdaBoost & 0.88 & 0.58 & 0.69 \\ \hline
		Random Forest & 0.79 & 0.71 & 0.74 \\ \hline
		
	\end{tabular}
	\label{tb:pr_rec_f1}
\end{table}

%\begin{table*}[t]\small
%	\centering
%	\caption{Comparison of the methods on labeled data, for the positive (extremist) and negative (non-extremist) classes.}%\vspace{-2mm}
%	\small
%	\begin{tabular}{|l|c|c|c|c|c|c|c|c|} 
%		\cline{1-9}
%		\textbf{Learner} & \multicolumn{2}{c|}{\textbf{Precision}} &
%		\multicolumn{2}{c|}{\textbf{Recall}} &
%		\multicolumn{2}{c|}{\textbf{F1-score}} & 
%		\multicolumn{1}{c|}{\textbf{Accuracy}} &
%		\multicolumn{1}{c|}{\textbf{AUC}}
%		\\
%		\hhline{=========}
%		& $class_p$ & $class_n$ & $class_p$ & $class_n$ & $class_p$ & $class_n$ & total & total\\ \cline{2-9}
%		Laplacian SVM & 0.89 & \textbf{0.74} & 0.6 & 0.95 & 0.7 & \textbf{0.81} & 0.77 & 0.75\\ \hline
%		LabelSpreading (RBF) & 0.85 & 0.73 & 0.69 & 0.87 & \textbf{0.76}  & 0.79  & \textbf{0.78} & \textbf{0.78}\\ \hline
%		LabelSpreading (KNN) & 0.83 & 0.71 &  0.67 & 0.84 & 0.73  & 0.76  & 0.75 & 0.75\\ \hline
%		Co-Training (SVM) & 0.9 & 0.66 & 0.53  & 0.95 & 0.66 & 0.73  & 0.73 & 0.74 \\ \hline
%		SVM & \textbf{0.96} & 0.66  & 0.5 & \textbf{0.98} & 0.65 & 0.79  & 0.74 & 0.74 \\ \hline
%		KNN & 0.81 & 0.72 & 0.7 & 0.82 & 0.74 & 0.76  & 0.76 & 0.75\\ \hline
%		Gaussian NB & 0.89 & 0.68 & 0.56 & 0.93 & 0.69 & 0.78  & 0.75 & 0.74\\ \hline
%		Logistic Regression & 0.76 & 0.68 & 0.61 & 0.82 & 0.67 & 0.73  & 0.71 & 0.71\\ \hline
%		AdaBoost & 0.88 & 0.68 & 0.58 & 0.92 & 0.69 & 0.78  & 0.75 & 0.75\\ \hline
%		Random Forest & 0.79 & 0.72 & \textbf{0.71} & 0.8 & 0.74 & 0.76 & 0.76 & 0.75\\ \hline
%	\end{tabular}
%	\label{tb:pr_rec_f1}
%\end{table*}

\begin{figure}
	\caption{Frequency of each handle-related feature in the filtered dataset for (left) labeled and (right) unlabeled instances. The most frequent features for labeled and unlabeled instances are \textit{Max \# of occurrence of a character in a handle} and \textit{Complexity of a handle}.}
	\centering
	\includegraphics[width=0.23\textwidth]{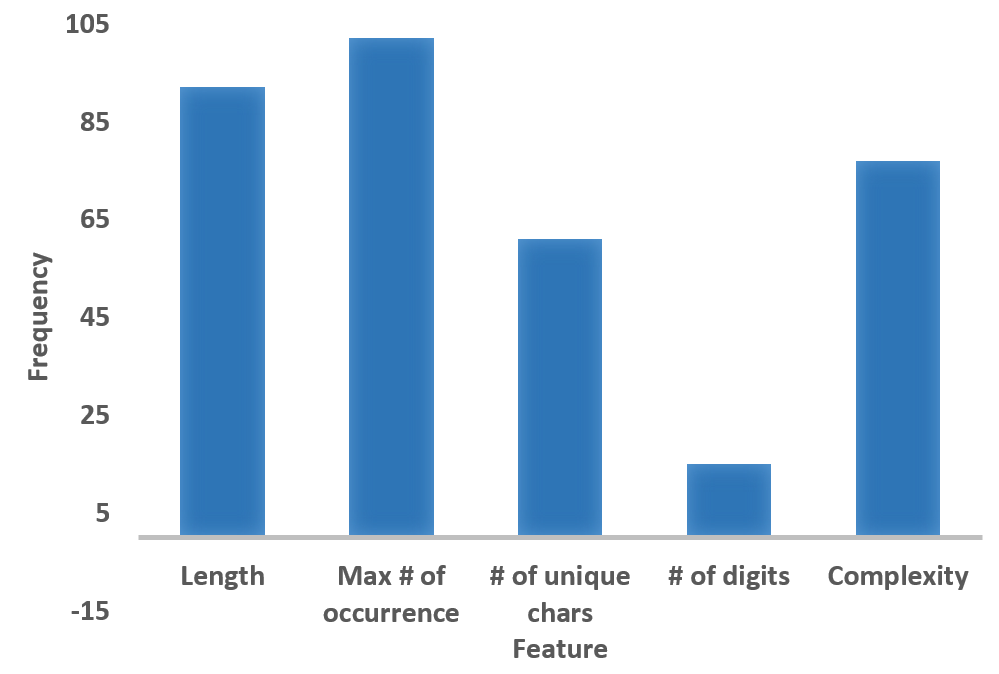}
	\includegraphics[width=0.23\textwidth]{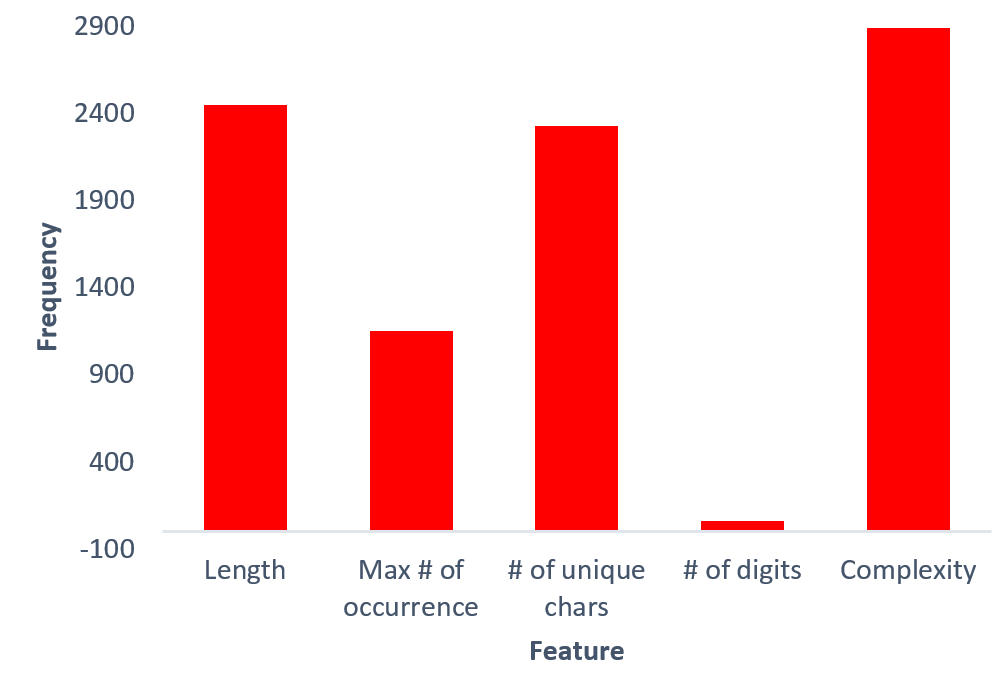}
	\label{fig:freq}
\end{figure}

\section{Experiments}
In this section, we first describe the learners used in this work and provide details on the parameters they use. Then, classification results are presented and finally significance of the features is discussed.

\subsection{Approaches}
%In this work, we present the results for the following learners:
\begin{itemize}
	\item \textbf{Semi-Supervised}: Laplacian support vector machines (SVM)~\cite{laplacian_svm}, graph inference-based label spreading approach~\cite{Zhou04learningwith} with radial basis function (RBF) and K-nearest neighbor (KNN) kernels, and co-training learner~\cite{blum1998combining} with two SVMs.
	\item \textbf{Supervised}: SVM, KNN, Gaussian na\"{i}ve Bayes, logistic regression, adaboost, random forest, and Char-LSTM~\cite{kim2016character}.
\end{itemize}

We note that supervised learners only use labeled instances for the training process, while semi-supervised algorithms use labeled and unlabeled instances~\cite{alvari2017semi}. 

For the sake of fair comparison, all algorithms were implemented and run in Python. Note for the methods that require special tuning of parameters, we performed grid search to choose the best set of parameters. Before going any further, we define the parameters used in each learner and then demonstrate their best picked values by our grid search.

\begin{itemize}
	\item \textbf{SVM: }Tolerance for stopping criteria was set to the default value of 0.001. Penalty parameter $C$ was set to 1 and linear kernel was used.
	\item \textbf{Char-LSTM: }This is similar to the character-aware models used for sequential word predictions. We adapt the neural network to a sequence classification problem where the inputs are the vector of one-hot encoding of each character of the handle and the output is the handle being classified as extremist or non-extremist. We set the maximum username length to 10, padding with zeros where necessary. We use an embedding layer after the input layer to convert each username dimension to 16. This is fed to a single layer LSTM module having 30 units.
	\item \textbf{LabelSpreading (RBF): }RBF Kernel was used and $\gamma$ was set to the default value of 20.
	\item \textbf{Laplacian SVM: }We used linear kernel and set the parameters   $C_l=0.6$  and   $C_s=0.6$.
	\item \textbf{LabelSpreading (KNN): }KNN kernel was used and the number of neighbors was set to 5.
	\item \textbf{Co-training (SVM): }We followed the algorithm introduced in~\cite{blum1998combining} and used two SVM as our classifiers. For both SVMs we set the tolerance for stopping criteria to 0.001 and the penalty parameter $C=1$.
	\item \textbf{KNN: }The number of neighbors was set to 5.
	\item \textbf{Gaussian NB: }There were no specific parameter to tune.
	\item \textbf{Logistic regression: }We used the $l2$ penalty. We also set the parameter $C=1$  (the inverse of regularization strength) and tolerance for stopping criteria to 0.01.
	\item \textbf{Adaboost: }The number of estimators was set to 200 and we also set the learning rate to 0.01.
	\item \textbf{Random forest: }We used 200 estimators and the entropy criterion was used.
\end{itemize}

\subsection{Classification Results}
We use tenfold cross-validation on the labeled data as follows. We first divided the set of labeled instances into 10 different sets of equal size. Each time, we held one set out for validation (we did this by removing their labels and adding them to the unlabeled instances). For the supervised learners, this set along with a set of the existing unlabeled samples are only used for the purpose of testing whereas for the semi-supervised setting, we use both sets in the training and testing phases. This procedure is performed for all approaches for the sake of fair comparison. Finally, we report the average of 10 different runs, using various evaluation metrics including precision, recall and F1-score in Table~\ref{tb:pr_rec_f1}. 

\textbf{Observations. }We make the following observations:

\begin{itemize}
	\item SVM achieve the highest precision of 0.96 in identifying online violent extremists, which shows the significance of the proposed feature set.	
	\item The semi-supervised LabelSpreading (RBF) was able to perform as good as Char-LSTM and they both achieve the highest F1-score on identification of extremists. This along with the fact that Char-LSTM has shown promising results in the literature while it does not use any of our hand-crafted features, further demonstrates the effectiveness of the introduced feature engineering scheme.
	\item For char-LSTM, we achieve a precision of 0.77 while maintaining a high recall of 0.76 on the positive class. This suggests that the memory module in LSTM can help in minimizing the number of false negatives.
\end{itemize}

Overall, the observations we make here suggest that the answer to the \textbf{second question} is positive-- using an existing set of labeled examples could help inferring the labels of unseen usernames.  

\begin{table}[!t]
	\centering
	\caption{Significance of the features using the labeled instances and $\chi^2$. The most significant feature for the labeled set is \textit{\# of unique characters in a handle}.
		}
	\begin{tabular}{|l|c|}
		\cline{1-2}
		\textbf{Feature} & $\chi^2$ \\
		\hhline{==}
		Length of the handle  & 22.73 \\ \hline
		Max \# of occurrence of a character  & 0.24  \\ \hline
		\# of unique characters  & \textbf{37.2} \\ \hline
		\# of digits the handle starts with  & 12.57 \\ \hline
		Complexity of the handle  & 3.18 \\ \hline
	\end{tabular}
	\label{tb:feat}
\end{table}	
\subsection{Significance of Features}
We conduct significance analysis of the features using the labeled instances and feature selection measure $\chi^2$. The results in Table~\ref{tb:feat} suggest that the most significant feature is the \textit{number of the unique characters in the username} while the least important one is the \textit{maximum number of occurrence of a character in the username}. This observation further demonstrates that frequency and importance of the features in the labeled dataset are not necessarily in line with each other and in fact are inversely related in our case. In other words, although \textit{maximum number of occurrence of a character in the username} is the most frequent feature in the labeled dataset, it is the least important feature in identification of online violent extremists according to the Fig.~\ref{fig:freq} where we depict the frequency of each feature for both labeled and unlabeled examples.

\section{Conclusion and Future Work}
In this work, we presented a scheme that using as little as three groups of information related to the Twitter handle, profile and textual content of users, can determine if a given handle could belong to an extremist. The framework first uses highly indicative patterns related to extremism to filter out less likely extremists. Ultimately, high likely extremist are identified using only features related to their usernames. 

In future, we would like to replicate the work by deploying more features and investigate if incorporating those features to the framework can lead to performance boost. We also plan to incorporate the feature space designed in this work into a semi-supervised learner as regularization terms in order to further increase the classification performance in detecting online violent extremists. Finally, since hand-labeling unlabeled examples is expensive, a valuable research direction would be to deploy active learning to enable iterative supervised learning to actively query for labels.  

\section{Acknowledgments}
This work was supported through DoD Minerva program.

%\bibliographystyle{IEEEtran}
%\bibliography{ref}

\end{document}